\documentclass[aps,twocolumn,groupedaddress]{revtex4}

\begin{document}
\draft
\title{Nonperturbative Dynamical Theory and A Scheme for Nonequilibrium Transport}
\author{Jongbae Hong and Wonmyung Woo}
\affiliation{Department of Physics and Astronomy \& Center for
Theoretical Physics, Seoul National University, Seoul 151-747,
Korea}

\date{\today}

\begin{abstract}
We develop a nonperturbative dynamical theory (NDT) that is useful
for treating nonequilibrium transport in a system with strong
correlation. We apply our NDT to the single-impurity Anderson
model in equilibrium to check its reliability by comparing with
the results of numerical renormalization group method (NRG). We
finally suggest a self-consistent loop to calculate the current in
a lead-dot-lead system with Kondo coupling.

\end{abstract}


\maketitle \narrowtext

One of the most problematic subjects in the field of theoretical
condensed matter physics is the treatment of nonequilibrium
transport\cite{keldysh} in a system with strong correlation. There
is no established theory for treating the interacting system when
it approaches nonequilibrium state, and a strongly correlated
system requires nonperturbative treatment. Moreover, methods based
on the static nonperturbative theory, such as
NRG\cite{nrg,Hewson}, are inappropriate for studying
nonequilibrium transport, because current is a dynamical quantity.
Therefore, a possible method for the treatment of nonequilibrium
transport in a system with strong correlation is by employing the
NDT, which has not been studied successfully thus far. Here, we
report an NDT that is developed by fully utilizing the dynamical
nature of the Heisenberg picture. We first apply the new NDT to
the single-impurity Anderson model in equilibrium; further, we
extend the application to the nonequilibrium transport of a
lead-dot-lead system.

Transport phenomena in mesoscopic systems have recently attracted
considerable interest in connection with nonequilibrium transport
in a strongly correlated system, which is one of challenging and
debated subjects in recent theoretical condensed matter physics.
An example of a typical quantum system with strong correlation
under nonequilibrium conditions is a quantum dot with metallic
leads under bias. Theoretical studies on quantum dot have usually
reported the conductance properties that can be obtained by using
the NRG, since conductance is an equilibrium property of the
system. Even though the NRG has been successful in providing
low-energy eigenvalues and eigenstates rigorously for a part of
strongly correlated systems, it cannot be a resolver of the
nonequilibrium transport problem.

A good formalism for nonequilibrium transport has been established
by Meir and Wingreen\cite{meir} in terms of nonequilibrium Green's
functions. Let us consider the motion of a spin-up electron in a
quantum dot under bias. Since the movement of a spin-up electron
will be affected by the movement of the spin-down electron due to
strong correlation at the dot, the retarded on-site Green's
function of a spin-up electron must contain information on the
back and forth movement of the spin-down electron. A static theory
such as NRG is incapable of yielding information on the back and
forth movement of an electron with a particular spin. A dynamical
theory, however, can incorporate this information in Green's
function. In this work, we present an NDT giving the
nonequilibrium Green's function that can provide a spectral
density for the Anderson model when the system is in equilibrium,
which is comparable to the one obtained by the NRG, and a scheme
providing the current-voltage characteristics under nonequilibrium
conditions.

The Heisenberg picture stresses the dynamics of operators compared
with other two pictures. However, the advantage of its dynamical
nature has not received sufficient appreciation. Here, we focus on
the dynamical nature of the Heisenberg picture to develop the NDT.
The formal solution of the Heisenberg equation for a fermion
annihilation operator $c_{d\uparrow}$, i.e.,
$c_{d\uparrow}(t)=c_{d\uparrow}+[{\hat
H},c_{d\uparrow}](it)+[{\hat H},[{\hat
H},c_{d\uparrow}]](it)^2/2+\cdots,$ where ${\hat H}$ is the
Hamiltonian of the system, describes the time evolution of
$c_{d\uparrow}$. The operators in each term of the expansion
represent linearly independent ways of annihilation of a spin-up
electron at site $d$ at time $t$. These operators provide linearly
independent bases spanning the operator or Liouville space. We
call these the dynamical bases. The inner product in the Liouville
space is defined as $\langle{\hat A}|{\hat B}\rangle\equiv
\langle\{{\hat A},{\hat B}^\dagger\}\rangle$, where ${\hat A}$ and
${\hat B}$ are elements of the Liouville space, the curly brackets
denote the anticommutator, and the angular brackets indicates an
equilibrium or nonequilibrium average, depending on the situation.
Constructing appropriate dynamical bases is the essence of the
NDT.

We first develop our NDT for the single-impurity Anderson model
represented by the Hamiltonian ${\hat
H}=\sum_\sigma\epsilon_dc^\dagger
_{d\sigma}c_{d\sigma}+\sum_{k,\sigma}\epsilon_kc^\dagger_{k\sigma}
c_{k\sigma}+\sum_{k,\sigma}(V_{kd}c^\dagger
_{d\sigma}c_{k\sigma}+V^*_{kd}c^\dagger_{k\sigma}c_{d\sigma})
+Un_{d\uparrow}n_{d\downarrow},$ where
$n_{d\downarrow}=c_{d\downarrow}^{\dagger} c_{d\downarrow}$, and
the subscripts $\sigma$, $k$, and $d$ denote the spin index,
quantum state of the metallic reservoir, and the position of
impurity, respectively, and then compare the result for the
spectral density with that of the NRG. Finally, we apply our NDT
to study the nonequilibrium transport of the lead-dot-lead system
with on-site Coulomb interaction.

The basis operators created by the commutators with the
Hamiltonian are composed of $c_{k\uparrow}$ where $k=d, 1, 2,
\cdots, \infty$ and the operators combined  with $c_{k\uparrow}$
and other operators such as $n_{d\downarrow}$,
$j^-_{d\downarrow}$, and $j^+_{d\downarrow}$, where
$j^-_{d\downarrow}=i[{\hat H},
n_{d\downarrow}]=i(\sum_kV_{kd}c^\dagger
_{k\downarrow}c_{d\downarrow}-\sum_kV^*_{kd}c^\dagger_{d\downarrow}c
_{k\downarrow})$, i.e., the current operator, and
$j^+_{d\downarrow}=i[n_{d\downarrow},
j^-_{d\downarrow}]=(\sum_kV_{kd}c^\dagger
_{k\downarrow}c_{d\downarrow}+\sum_kV^*_{kd}c^\dagger_{d\downarrow}c
_{k\downarrow})$. We keep the meaningful operators that describe
the virtual exchange between spin-up and spin-down electron and
take the mean-field approximation for other part of the primitive
basis operator. After this manipulation, appropriate linearly
independent dynamical bases spanning a reduced Liouville space of
the operator $c_{d\uparrow}$ can be constructed. The bases except
$c_{d\uparrow}$ itself must be orthogonal to $c_{d\uparrow}$ in
order to yield a correct projection $
\langle\{c_{d\uparrow}^\dagger,c_{d\uparrow}(t)\}\rangle$ that
gives the on-site retarded Green's function.

The dynamical bases of the reduced Liouville space of
$c_{d\uparrow}$ are composed of five parts. The first two are (i)
a set of bases $S_k$ defined by $S_k=\{c_{k\uparrow}|k=1, 2,
\cdots, \infty\}$ for describing the annihilation at the impurity
site after some hoppings in the metal and (ii) a set of bases
$S_n$ defined by $S_n=\{\delta n_{d\downarrow}c_{k\uparrow}|k=1,
2, \cdots, \infty\}$ where $\delta
n_{d\downarrow}=n_{d\downarrow}-\langle n_{d\downarrow}\rangle$
for describing the number fluctuation of the spin-down electron at
the impurity site during the annihilation process of (i). Lastly,
we consider the bases coupled to $c_{d\uparrow}$, i.e., (iii) a
set of bases $S_d $ defined by $S_d=\{c_{d\uparrow}, \delta
j^-_{d\downarrow}c_{d\uparrow}, \delta
j^+_{d\downarrow}c_{d\uparrow} \}$ for describing the annihilation
of a spin-up electron at the impurity site without any coupled
processes and with the processes coupled to the back and forth
fluctuation of the spin-down electron at the impurity site. The
latter describes the hybridization in the Kondo process.

\begin{figure}[t]
\vspace*{5cm} \includegraphics{fig1.eps} \vspace*{0.5cm} \caption{
Pictorial descriptions of the roles of basis operators
$n_{d\downarrow}c_{k\uparrow}$ (a) and
$j^\mp_{d\downarrow}c_{d\uparrow}$ for a particular $\bf k$-state
(b). Sum over all $\bf k$ in (b) becomes
$j^\mp_{d\downarrow}c_{d\uparrow}$. The dashed arrows do not
indicate hybridization. They just indicate creation and
annihilation.}
\end{figure}

The bases of the first subset $S_k$ and $c_{d\uparrow}$ constitute
a complete set when the system is noninteracting. Since $\delta
n_{d\downarrow}^2$ is composed of a constant and $\delta
n_{d\downarrow}$, the first two subsets, $S_k$ and $S_n$, along
with $c_{d\uparrow}$, contain all the dynamical processes
described by number fluctuations of a spin-down electron during
the annihilation process of a spin-up electron at site $d$. The
operators in the third subset, $\delta
j^\pm_{d\downarrow}c_{d\uparrow}$, contribute to describing the
Kondo effect. As an example, the pictorial descriptions of the
basis operators $\delta n_{d\downarrow}c_{k\uparrow}$  and $\delta
j^\pm_{d\downarrow}c_{d\uparrow}$ are shown in Fig. 1. We do not
consider the other complicated processes that can occur during the
annihilation process of a spin-up electron at time $t$ since the
processes considered above may be sufficient for describing the
Kondo phenomena in the single-impurity Anderson model. In
addition, the operator $c_{d\uparrow}$ is orthogonal to all other
bases of the reduced Liouville space.

We are now in a position to obtain the spectral density of a
spin-up electron at site $d$ by NDT using the dynamical bases
introduced above. The resolvent Green's function operator in the
Heisenberg picture is written as ${\hat G}^{\pm}=(\omega\pm
i\eta-{\rm\bf L})^{-1}$, where ${\rm\bf L}$ is the Liouville
operator defined by ${\rm\bf L}{\hat A}={\hat H}{\hat A}-{\hat
A}{\hat H}$, $\eta$ is a positive infinitesimal, and the
superscripts $\pm$ denote retarded ($+$) and advanced ($-$),
respectively. This expression becomes ${\hat G}^{\pm}=(\omega\pm
i\eta-{\hat H})^{-1}$ in the Schr\"{o}dinger picture.

Since the retarded and advanced Green's functions are given by
$G^{\pm}_{ij}(\omega)=\langle{\hat e}_i|(\omega\pm i\eta-{\rm\bf
L})^{-1}|{\hat e}_j\rangle,$ where $|{\hat e}_j\rangle$ is one of
basis operators spanning the Liouville space, they are given by
the inverse of the matrix ${\rm\bf M}$ defined by ${\rm\bf
M}_{ji}=\langle{\hat e}_i|z{\rm\bf I}+i{\rm\bf L}|{\hat
e}_j\rangle$, where $z=-i\omega\pm\eta$, i.e.,
$iG^{\pm}_{ij}(\omega)=\langle{\hat e}_i|{\rm\bf M} ^{-1}|{\hat
e}_j\rangle=({\rm adj}\,\, {\rm\bf M})_{ij}[{\rm det} \,\,{\rm\bf
M}]^{-1}$, where $({\rm adj}\,\, {\rm\bf M})_{ij}$ denotes the
cofactor of the $ji$-element in the determinant of ${\rm\bf
M}$\cite{Hong}. This expression has also been reported in
literature\cite{fulde}.

In constructing matrix ${\rm\bf M}$, we use normalized bases in
order to make it quasi-symmetrical. If we arrange $S_k, S_n$, and
$S_d$ in a regular sequence to construct the matrix ${\rm\bf M}$
for the single-impurity Anderson model, it is represented by a
matrix of four blocks,
\[ {\rm\bf M}_{{\rm SIAM}}=\left( \begin{array}{cc} \,\, \,\, {\rm\bf M}_{r} \,\,
{\rm\bf M}_{dr} \\ -{\rm\bf M}_{dr}^*  \,\, {\rm\bf M}_{d}
\end{array} \right),\] where the subindex $r$ represents metallic reservoir.
The blocks ${\rm\bf M}_{r}$, ${\rm\bf M}_{dr}$, and ${\rm\bf
M}_{d}$ are $\infty\times\infty$, $3\times\infty$, and $3\times 3$
matrices, respectively. Since the inner product has a relation
$\langle i{\rm\bf L}{\hat A}|{\hat B}\rangle=-\langle i{\rm\bf
L}{\hat B}|{\hat A}\rangle^*$, only the real parts of the matrix
elements have different signs for their counter-parts.

The block ${\rm\bf M}_{r}$ is represented by
\[ {\rm\bf M}_{r}=\left[ \begin{array}{cc} {\rm\bf M}_{11} \, \, \, {\bf 0}\, \, \\
\, \,\, \,  {\bf 0} \, \, \, \, \,   {\rm\bf M}_{11}
\end{array} \right], \] where ${\rm\bf M}_{11}$ is diagonal and its
elements are $z+i\epsilon_{k}$, $k=1, 2, \cdots, \infty$. The
$3\times\infty$ block ${\rm\bf M}_{dr}$, on the other hand, has
the form
$${\rm\bf M}_{dr}=\left[ \begin{array}{ccc} {\rm\bf C_{kd}} \, \, \,\,\,
{\rm\bf 0} \,\, \, \, \, \,\, \, \, \, \, \,\, \, {\rm\bf 0}\, \,
\, \, \\ \, \, \, \, \, {\rm\bf 0} \,\,  \, \,  \, \, \, \,
{\rm\bf C_{nj^-}} \,\, \, \, {\rm\bf C_{nj^+}}
\end{array} \right],
$$
where the column ${\rm\bf C_{kd}}$ has $iV_{kd}$ as its elements,
while ${\rm\bf C_{nj^-}}$ and ${\rm\bf C_{nj^+}}$ have $\xi_d^-
V_{kd}$ and $\xi_d^+ V_{kd}$, respectively. The blocks except
${\rm\bf M}_{d}$ contribute to the self-energy. The 11-element of
the block ${\rm\bf M}_{d}$ is $z+i(\epsilon_d+\langle
n_{d\downarrow}\rangle U)$, and the other diagonal elements are
given by $z+i\epsilon_d+[U\langle n_{d\downarrow}\delta j^{\mp
2}_{d\downarrow}\rangle/\langle\delta j^{\mp
2}_{d\downarrow}\rangle]$. These are equal to the 11-element of
${\rm\bf M}_{d}$ under the decoupling approximation. The
off-diagonal elements of ${\rm\bf M}_{d}$, on the other hand, are
given by ${\rm M}_{d}^{12}=-{\rm M}_{d}^{21}=(U/2)\xi_d^-$, ${\rm
M}_{d}^{13}=-{\rm M}_{d}^{31}=(U/2)\xi_d^+$, and ${\rm
M}_{d}^{23}=-{\rm M}_{d}^{32}=\gamma$, where
$2\xi_d^\mp=[i(1-2\langle n_{d\downarrow}\rangle)\langle
j^\mp_{d\downarrow}\rangle+\langle
i[n_{d\downarrow},j^\mp_{d\downarrow}] (1-2n_{d\uparrow})\rangle]
/[\langle (\delta j^\mp_{d\downarrow})^2\rangle^{1/2}\langle
(\delta n_{d\downarrow})^2\rangle^{1/2}]$ and
$\gamma=-2i\sum_kV_{kd}^* \langle
j^-_{d\downarrow}j^+_{d\downarrow}
c_{k\uparrow}c^\dagger_{d\uparrow}\rangle/[\langle (\delta
j^-_{d\downarrow})^2\rangle\langle (\delta
j^+_{d\downarrow})^2\rangle]^{1/2}$. These factors will be
determined when we calculate the spectral densities.

In order to handle the infinite dimensional matrix, matrix
reduction by L\"{o}wdin's partitioning technique\cite{loewdin} is
performed by solving the eigenvalue equation for the original
matrix ${\rm\bf M}_{{\rm SIAM}}$, such as ${\rm\bf M}_{\rm
SIAM}{\rm\bf C}={\rm\bf 0}$, where ${\rm\bf C}$ and ${\rm\bf 0}$
are infinite dimensional column vectors. The column vector
${\rm\bf C}$ is partitioned into two parts, i.e., infinite
dimensional ${\rm\bf C}_{r}$ and three dimensional ${\rm\bf
C}_{d}$, symbolizing the reservoir and impurity parts,
respectively. Then, the equation for ${\rm\bf C}_{d}$ is obtained
as $({\rm\bf M}_{d}-{\rm\bf M}_{rd}{\rm\bf M}_{d}^{-1}{\rm\bf
M}_{dr}) {\rm\bf C}_{d}\equiv {\widetilde {\rm\bf M}}_{d}{\rm\bf
C}_{d}={\rm\bf 0}$. The reduced $3\times 3$ matrix
${\widetilde{\rm\bf M}}_{d}$ contains the information on the
many-body dynamics of a spin-up electron starting from the
impurity site at $t=0$ and ending at it at time $t$.

It is impossible to calculate the inverse of a general
$\infty\times\infty$ matrix; however, this is not the case for the
matrix ${\rm\bf M}_{r}$, which is block diagonal. The inverse of
${\rm\bf M}_{r}$ can be obtained in a straightforward
manner\cite{house}. The second term appears as additional
self-energy terms in ${\widetilde{\rm\bf M}}_{d}$ after reduction.
One can easily imagine that $\xi_d^-=\xi_d^+$ in the Kondo regime
at half-filling, and the final form of ${\widetilde{\rm\bf
M}}_{d}$ for a symmetric Anderson model in which $\epsilon_d=-U/2$
and $\langle n_d\rangle=1/2$ is given by
\[ \widetilde{\rm\bf M}_{d}=\left( \begin{array}{ccc} -i\omega+i\Sigma_0
\,\,\,\,\,\,\,\, \,\,\, \,\,\,\, \, \,\, U\xi_d^-/2 \,\,\,\, \,\,
\,\,
\,\, \,\,\,\, \,\, \,\, \,\, U\xi_d^-/2 \,\, \,\,\,\, \\ \\
-U\xi_d^-/2 \,\,\,\,\,\, \,\, \,\, \,\,\,
-i\omega+i\xi_d^{-2}\Sigma_0 \,\, \,\,\, \,\, \,\, \,\,\,\,
\gamma+i\xi_d^{-2}\Sigma_0
\\ \\ \,\, -U\xi_d^-/2 \,\, \,\,\, \,\, \,\, \,\, \,\,
-\gamma+i\xi_d^{-2}\Sigma_0 \,\, \,\, \,\,\,
-i\omega+i\xi_d^{-2}\Sigma_0
\end{array} \right), \]
where $\Sigma_0=\sum_{\bf k}|V_{kd}|^2/(\omega-\epsilon_{\bf
k}+i\eta)\equiv\Lambda(\omega)-i\Delta(\omega)$ is the self-energy
of the Anderson model with $U=0$. We will neglect the ${\bf k}$
dependence of $V_{kd}$ in this work.  The real and imaginary parts
are respectively $\Lambda(\omega)=({\rm
P}/\pi)\int_{-\infty}^{\infty}
\Delta(\epsilon)d\epsilon/(\omega-\epsilon)$, where ${\rm P}$
represents the principal integration  and
$\Delta(\omega)=\pi\sum_k|V_{kd}|^2\delta(\omega-\epsilon_{\bf
k})=\Delta[1-(\omega/D)^2]^{1/2}$ for the semielliptical band,
where $\Delta=2|V|^2/D$ and $D$ is half of the band width. Now,
one can obtain the retarded Green's function that is given by
$iG^+_{dd}(\omega)=({\rm adj}\,\, \widetilde{\rm\bf
M}_{d})_{11}[{\rm det} \,\,{\widetilde{\rm\bf M}_{d}}]^{-1}$ by
calculating the inverse of matrix $\widetilde{\rm\bf M}_{d}$.

\begin{figure}[t]
\vspace*{6cm} \includegraphics{fig2.eps} \vspace*{0.2cm} \caption{
Spectral densities of the symmetric Anderson model with various
Coulomb repulsions.  }
\end{figure}

We find that $\gamma$ determines the width of the Kondo peak,
while $\xi_d^-$ governs the spacing of the side-peaks of the
spectral density. Therefore, from the analysis at atomic limit,
one can find that $\xi_d^-=1/\sqrt{2}$. On the other hand, our
retarded Green's function gives the real and imaginary parts of
the self-energy as ${\rm Re}\Sigma^+_{\uparrow
U}(\omega)=-(U^2/4\gamma^2)\omega+O(\omega^3)$ and $(1/\Delta){\rm
Im}\Sigma^+_{\uparrow
U}(\omega)=(U^2/2\gamma^4)\omega^2+O(\omega^4)$. The subscript $U$
indicates the interaction part of the self-energy. The real part
of the self-energy provides the wavefunction renormalization as
$Z=1/[1+(U^2/4\gamma^2)]$.

Unfortunately, it is difficult to obtain $\gamma$ by direct
calculation. Therefore, we determine rigorous $\gamma$ by using
the result of Bethe ansatz, which is given by
$Z^{BA}=(4/\pi)\sqrt{U/\Gamma}{\rm exp}[-\pi
U/4\Gamma+\pi\Gamma/4U]$ for the symmetric Anderson model[3]. We
plot the spectral density $\rho_{d\uparrow}(\omega)=-(1/\pi){\rm
Im} G^+_{dd}(\omega)$ in Fig. 2. Our result naturally compatible
with that of NRG at least in the Kondo regime.

The above result can be extended to nonequilibrium transport in a
lead-dot-lead system with on-site Coulomb interaction at the dot
in a straightforward manner because it is simply a single-impurity
Anderson model with two separate metallic reservoirs. This only
necessitates the simple requirement of an additional number of
bases for describing the left and right leads and the movements
from or to both leads, such as $S_k^L$, $S_n^L$ for the left lead,
$S_d^{\ell d\ell}\equiv\{\delta j^{-L}_{d\downarrow}c_{d\uparrow},
\delta j^{+L}_{d\downarrow}c_{d\uparrow}, c_{d\uparrow}, \delta
j^{-R}_{d\downarrow}c_{d\uparrow}, \delta
j^{+R}_{d\downarrow}c_{d\uparrow}\}$ for the dot, and $S_k^R$,
$S_n^R$ for the right lead. The superscripts $L$ and $R$ denote
the left and right leads, respectively.

If we arrange $S_k^L$, $S_n^L$, $S_d^{\ell d\ell}$, $S_k^R$,
$S_n^R$ in a regular sequence to construct the matrix ${\rm\bf M}$
for a lead-dot-lead system, it will be represented by a matrix of
nine blocks,
\[ {\rm\bf M}_{\ell d\ell}=\left( \begin{array}{ccc} {\rm\bf M}_{LL} \,\,\,\,\,
{\rm\bf M}_{dL} \,\, \,\, \,\,{\rm\bf 0}\,\,\,\,\,\,\, \\ {\rm\bf
M}_{Ld} \,\,\,\,\,\, {\rm\bf M}_{d} \,\,\,\,\,\,\, {\rm\bf M}_{Rd}
\\\,\,\,\, {\rm\bf 0} \,\,\,\,\,\,\,\,\,\,\,\, {\rm\bf M}_{dR}
\,\,\,\, {\rm\bf M}_{RR}
\end{array} \right),
\] where blocks ${\rm M}_{d}$, ${\rm M}_{dL}$ and ${\rm M}_{dR}$, and
${\rm M}_{Ld}$ and ${\rm M}_{Rd}$ are $5\times 5$,
$5\times\infty$, and $\infty\times 5$ matrices, respectively.
Blocks ${\rm\bf M}_{LL}$ and ${\rm\bf M}_{RR}$ are
$\infty\times\infty$ matrices that are constructed by the sets of
bases describing left and right leads, respectively. Since no
direct coupling exists between the left and right leads, zero
matrices are present at two of the corners. The structure of each
block is similar to that of the corresponding block of the matrix
${\rm\bf M}_{{\rm SIAM}}$.

The matrix reduction produces a $5\times 5$ matrix for
$\widetilde{\rm\bf M}_{d}$, whose inverse yields the on-site
retarded Green's function $G_{dd}^+(\omega)$ of the lead-dot-lead
system in the framework of the NDT, which is given by a function
of $\langle n_{d\sigma}\rangle$, $\langle
j^{-L,R}_{d\sigma}\rangle$, $\langle j^{+L,R}_{d\sigma}\rangle$.
Since the current is expressed by
$J_\sigma=J_\sigma^L=-J_\sigma^R=(J_\sigma^L-J_\sigma^R)/2$, where
$J_\sigma^L=(e/\hbar)\langle j^{-L}_{d\sigma}\rangle$, the current
and other variables are expressed by the lesser and retarded
Green's functions such that\cite{keldysh, meir}
\begin{eqnarray} \langle
j^{-L}_{d\sigma}\rangle&=&\int\frac{d\omega}{\pi}
\left\{i{\Gamma}_{\sigma}^L(\omega)G_{dd\sigma}^<(\omega)\right.
\nonumber \\
&-&\left. 2f_{L}(\omega)\Gamma_{\sigma}^L(\omega){\rm
Im}G_{dd\sigma}^+(\omega)\right\}\end{eqnarray}
\begin{equation}
\langle j^{+L}_{d\sigma}\rangle=
\int\frac{d\omega}{\pi}[f_{L}(\omega){\Gamma}_{\sigma}^{L}(\omega)
{\rm
Re}{G}_{dd\sigma}^+(\omega)-i{\rm}{G}_{dd\sigma}^<(\omega)\Lambda(\omega)],
\end{equation}
\begin{equation}
\langle n_{d\sigma}\rangle=\frac{-i}{2\pi}\int\! d\omega
G_{dd\sigma}^<(\omega)=\frac{-1}{\pi}\int\! d\omega
\tilde{f}(\omega){\rm Im}G_{dd\sigma}^+(\omega),\end{equation}
 where the effective Fermi
distribution $\tilde{f}(\omega)$ is given by\cite{oguri}
\begin{equation}\tilde{f}(\omega)=\frac{f_L(\omega){\Gamma}_\sigma^L(\omega)
+f_R(\omega){\Gamma}_\sigma^R(\omega) +\frac{1}{i}\Sigma^<_{\sigma
U}(\omega)}
{{\Gamma}_\sigma^L(\omega)+{\Gamma}_\sigma^R(\omega)-2{\rm
Im}\Sigma^+_{\sigma U}(\omega)}.\end{equation}

In order to construct a self-consistent loop for calculating the
current-voltage characteristics, we use the Keldysh
equation\cite{keldysh}
\begin{equation}G_\sigma^<(\omega)=G_\sigma^+(\omega)
\Sigma_\sigma^<(\omega)G_\sigma^-(\omega).\end{equation} Then, one
can construct a self-consistent loop from Eqs. (1)-(5) as

$\langle n_{d\sigma}\rangle^{(0)}, \langle
j^\mp_{d\sigma}\rangle^{(0)} \rightarrow
{G}_{dd\sigma}^{+(0)}(\omega)\rightarrow
\tilde{f}^{(0)}_\sigma(\omega)\rightarrow
G_{dd\sigma}^{<(0)}(\omega)\hspace{1cm}\\$ $\rightarrow \langle
n_{d\sigma}\rangle^{(1)}, \langle
j^\mp_{d\sigma}\rangle^{(1)}\rightarrow
{G}_{dd\sigma}^{+(1)}(\omega)\rightarrow\Sigma_{\sigma
U}^{<(1)}(\omega)\rightarrow \tilde{f}^{(1)}_\sigma(\omega)\\
\rightarrow \cdots.$

\noindent This iterative method for calculating nonequilibrium
quantities of the strongly correlated system is the second major
result of this work. The specific results will be reported in a
separate work.

This work was supported by Korea Research Foundation Grant No.
KRF-2003-070-C00020.

\end{document}